\documentclass[12pt]{article}

\usepackage{epsfig}
\renewcommand{\theequation}{\thesection.\arabic{equation}}
\makeatletter
\@addtoreset{equation}{section}
\makeatother

\newlength{\dinwidth}                       
\newlength{\dinmargin}                      
\begin{document}
\def\beq{\begin{equation}}
\def\eeq{\end{equation}}
\def\beqar{\begin{eqnarray}}
\def\eeqar{\end{eqnarray}}
\def\barr#1{\begin{array}{#1}}
\def\earr{\end{array}}
\def\bfi{\begin{figure}}
\def\efi{\end{figure}}
\def\btab{\begin{table}}
\def\etab{\end{table}}
\def\bce{\begin{center}}
\def\ece{\end{center}}
\def\nn{\nonumber}
\def\disp{\displaystyle}
\def\text{\textstyle}
\def\fs{\footnotesize}
\def\arraystretch{1.2}

\thispagestyle{empty}
\def\thefootnote{\fnsymbol{footnote}}
\setcounter{footnote}{1}

\hfill \begin{minipage}[t]{4cm}BI-TP 96/33\\
GSI - 96 - 43\\
September 1996  
\end{minipage}

\vspace*{2cm}
\begin{center}
{\Large\bf Thermal Dileptons from $\pi - \rho$ Interactions }   \\
{\Large\bf in a Hot Pion Gas}
\end{center}

\vspace*{0.3cm}
\begin{center}
{\bf R.~Baier}$^1$, {\bf M.~Dirks}$^1$  {\bf and K.~Redlich}{$^{2,}$}{$^3$} 
\\[.3cm]
$^1${\it Fakult\"at f\"ur Physik, Universit\"at Bielefeld, 
D-33501 Bielefeld, Germany}\\
$^2${\it Institute for Theoretical Physics, University of Wroclaw, \\
PL-50204 Wroclaw, Poland}\\
$^3${\it GSI, PF 110552,  D-64220 Darmstadt, Germany}
\end{center}

\vspace*{3cm}
\section*{Abstract}
A systematic study of low mass  dilepton  production  
 from    $\pi -\rho $ interactions
in a hot medium is presented. 
Applying  finite temperature perturbation theory
 the dilepton rate, respectively the virtual photon 
 rate, is  computed up to order $g_\rho^2$. 
For dilepton masses below the $\rho$ 
 the two-body  reactions 
 $\pi\pi\to \rho \gamma^*$,  $\pi\rho\to \pi \gamma^*$,
and the decay process $\rho \to \pi \pi \gamma^*$ 
give significant contributions. 
Non-equilibrium contributions to the thermal rate are estimated,
including the modification of the  particle  distribution function
with non-zero pion chemical potential.
The comparison of the dilepton rate with the recent
data measured in nucleus-nucleus collisions at SPS energy 
by the CERES Collaboration
is also performed. It is shown that the additional thermal dileptons
from $\pi-\rho $ interactions can partially account for the access
 of the soft dilepton yield seen experimentally.

\vfill

\def\thefootnote{\arabic{footnote}}
\setcounter{footnote}{0}
\clearpage

\section{Introduction}

\bigskip 
Dilepton  production 
is  one of the interesting tools 
to study  collective effects in strongly interacting matter produced in 
ultrarelativistic heavy ion collisions \cite{shur,mclerran,ruuskanen,redl}.
This is particularly 
evident from the recent CERN experimental data   at 
 SPS  collision energy. A significant enhancement 
 of the      dilepton spectrum measured in A-A 
 collisions as compared with 
 p-p and p-A collisions is reported. 
Two experiments, CERES/NA45 \cite{ceres}   and HELIOS/3 \cite{helios} 
 have measured dileptons 
in the low mass range in S-Au, Pb-Au  and S-W collisions respectively. 
Both  these experiments have reported the excess of dileptons 
with invariant masses between $0.25$  and  $ 1$ GeV 
in A-A nuclear collisions, compared to 
 expected \cite{ceres} or 
 measured \cite{helios} 
 spectra  in p-A  collisions. 

These  remarkable results indicate the appearance of  collective 
phenomena in heavy ion collisions, 
like  thermalization of the medium  after the 
collision. 
Since the temperatures are not yet to be expected as high,
in particular the reactions involving   pions and rho 
 mesons are important and have to be considered. 
In the kinematical window of low mass dileptons  pion annihilation 
        with the $\rho$ in the intermediate state  is a basic 
 source 
of thermal dileptons, also responsible  
for the low mass  dilepton enhancement reported by the CERN
experiments. 
Indeed, recent theoretical calculations 
show that close to the $\rho$ peak  the thermal production rate 
due to  pion annihilation         is      compatible with the experimental 
data \cite{ko,dinesh}. However, neither the size 
  of the access below the $\rho$ peak 
nor the shape of the distribution is quantitatively explained by
this contribution alone. 
In a thermal medium, however, the partial restoration of chiral 
symmetry in hot and dense hadronic matter may modify 
  the properties of vector mesons 
especially their mass or decay width \cite{gery,koch}: 
in  Ref.~\cite{ko} it is shown 
 that thermal dilepton production due to the pion 
 annihilation  process together with the assumption 
of a decreasing rho meson mass leads to a quantitative explanation 
of the  enhancement of the low-mass dileptons observed by the CERES 
and by the HELIOS/3 experiments. 

The above results together with previous 
 theoretical studies \cite{gale,gale2,song} 
 suggest the  importance of pion scattering in a medium 
as a source of soft dileptons. 
 In a thermal medium, however, $\pi^+\pi^-\to e^+e^-$ is not the only 
 process which has to be considered.
 For example the contribution  from the two-body reaction 
 $\pi^+\pi^-\to \rho \gamma^* \to \rho e^+e^- $ 
 does not have a kinematical threshold at $2 m_\pi$, and
 therefore certainly dominates the 
 basic pion annihilation process 
 for dilepton masses  $M\simeq 2m_\pi$.
 This example indicates, that a more  complete 
 analysis of the low mass dilepton 
spectrum originating from a possible thermal medium, requires further 
considerations.

 Previous calculations of the thermal rate of emission of 
direct, i.e. real, photons with energies less than 
$1$ GeV have  shown that there are two 
 contributing reactions involving a neutral $\rho$: 
 the annihilation process  
 $\pi^+\pi^-\to \rho^0 \gamma$ and 
Compton like scattering $\pi^{\pm} \rho^0 \to \pi^{\pm} \gamma$ \cite{kapusta}.
Both contribute to dilepton 
production, where  the real photon is replaced by a virtual one.
The importance of two-body processes for low mass dileptons
has also been stressed in the case of production from
the quark-gluon plasma \cite{pisarski}.

 In this paper we  compute the contributions to 
 the dilepton production rate 
from $\pi-\rho$ interactions for invariant masses 
up to the $\rho$ peak.
In sect.~2, after shortly treating the $\pi^+ \pi^- \to \gamma^*$
Born rate, we discuss  
the "real" $2\to 2$  and $\rho \to \pi \pi \gamma^*$ reactions,
at the two-loop level of the virtual photon selfenergy.
Also the "virtual" two-loop corrections to the Born rate
are derived.
Sect.~3 extends the estimates of the dilepton rate taking into account
out of equilibrium effects by introducing the pion chemical potential.
Finally, in sect.~4 we compare the dilepton rate with 
experimental data \cite{ceres} in the low mass region.

\bigskip 

\section{Dilepton production and $\pi -\rho$ interactions }\label{secdipro}

The thermal dilepton production rate is considered in lowest order of
the electromagnetic coupling. First, the medium is taken at rest.
The rate per unit space-time volume, $dR \equiv dN/d^4x$,
is related to the absorptive part of the  
 photon self-energy tensor $\Pi_{\mu\nu}$ 
 \cite{mclerran,gale,weldon,weldon1}. 
For lepton pairs of invariant mass $M$, energy $q_0$ and momentum 
$\vec q$ the relation is  
\beq
{{dR}\over {dM^2{{d^3q}/ {q_0}}}}= -
{{\alpha}\over {24\pi^4 M^2}} n( q_0 ){\rm Im} 
\Pi^\mu_{~\mu}(q_0,\vec q )  ,
\label{2.2} 
\eeq 
where $n$ is the Bose distribution function
at temperature $T = 1/\beta$ (cf. Appendix A). 
The tensor is constrained by 
 current conservation $q_\mu\Pi^\mu_{~\nu} (q) =0$.  

We adopt the real-time formulation of finite temperature field theory
to evaluate ${\rm Im} \Pi^\mu_\mu$. Using the Keldysh variant
(with time path-parameter $\sigma = 0$ \cite{Lands,chou,Eijck})
the following 
useful relation holds in terms of the photon-selfenergy
 matrix $\Pi^{\gamma}_{ab}$, 
\beq
{\rm Im}\Pi^\mu_{~\mu} (q_0, \vec q) =  {{1}\over 
{2 n(q_0)}} \, i\Pi^{\gamma}_{12}(q_0,\vec q) .
\label{2.3}
\eeq 

\subsection{Born rate}

In order to  clarify the notation
we give the one-loop expression for $\Pi^{\gamma}_{12}$.
With the momentum labels indicated in Fig.~1 this pion-loop
contribution with the temperature-dependent propagators (\ref{A.1}),
summarized in Appendix A, reads
\beq
i \Pi^{\gamma}_{12}(q_0,\vec q) = - (-ie)^2 \int ~{{d^4p}\over {(2 \pi)^4}}
   (p + p')^2~ iD_{12} (p) iD_{21}(p') ,
\label{2.3a}
\eeq
where $p' = p - q$.

\begin{figure}[ht]
\centering
\epsfig{file=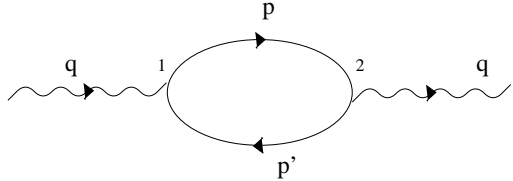, angle=-90, height=2.5 cm}
\caption{\label{rho}\it{One-loop photon selfenergy diagram.
The solid line denotes the pion}}
\end{figure}

Using the equilibrium distributions one recovers the standard
expression for the Born term
$\pi^+ \pi^- \to \gamma^* \to l^+~l^-$.
For masses $M$ near the $\rho$
it is appropriate \cite{gale} to implement VDM and to insert the
pion electromagnetic formfactor $F_\pi$,
\beq
|F_\pi(M)|^2 = {{m_\rho^4}\over {(M^2-{m_\rho^*}^2)^2 
 +\Gamma_\rho^2m_\rho^2}} , 
\label{2.5} 
\eeq 
into the production rate. For our  numerical
analysis the parameters are chosen as 
$m_\rho=0.775$ GeV, 
$m_\rho^*=0.761$ GeV and $\Gamma_\rho= 0.118$ GeV  
in agreement with the measured pion electromagnetic 
form factor \cite{gale}.
We do not consider the possibility of medium dependent
resonance parameters, as done in \cite{ko}. 
Also contributions due to the $A1$ resonance
\cite{Haglin} are neglected.

 For  soft dileptons with  invariant masses $M< m_\rho$ the  
 mass distribution per unit space time volume is obtained 
 within  a   very good approximation from the rate
with the heavy photon taken at rest with respect to the medium
(also kept at rest)
\cite{ruuskanen,koch}
\beq 
{{dN}\over {dM^2d^4x}}\simeq 
4\pi T^2 ({{\pi M}\over {2T}})^{1/2} 
{{dR}\over {dM^2{{d^3q}/ {q_0}}}}|_{\vec q=0} , 
\label{2.6} 
\eeq
where the Born rate for $M \simeq m_\rho$, 
 with the Boltzmann approximation of the pion distribution,
has the following simple form,
\beq 
{{dR^{Born}}\over {dM^2{{d^3q}/ {q_0}}}}|_{\vec q=0}= 
{{\alpha^2}\over {96\pi^4}} |F_\pi(M)|^2 \exp (-M/T) . 
\label{2.4} 
\eeq 
The more complete expression for the Born contribution
with quantum statistics and arbitrary heavy photon momentum
is, however, also well known in the literature (see e.g.~\cite{redl}).
It will be used in the following numerical estimates.
Throughout the paper we neglect the pion mass
(except for the numerical analysis of the Born rate
for $M < m_\rho$), and the masses
of the leptons.
Because of the many uncertainties we believe that these approximations
are very well justified.

Taking into account out of equilibrium effects in determining
the rates, the expression (\ref{2.4}) 
is multiplied by a factor 
\beq
 1 + 2 {\delta \lambda}  ,
\label{2.4a}
\eeq
with $\delta \lambda = \lambda - 1$  
in terms of the fugacity  $\lambda$ (cf. Appendix A). 
This first approximation, i.e. staying near equilibrium
for $\delta \lambda \ll 1$, and neglecting more complicated
dependences,  should indicate
the possible  increase ($\lambda > 1$) or the decrease ($\lambda < 1$) of the
dilepton rate due to non-equilibrium effects. 

\subsection{Two-loop contributions}

    The interaction of the charged pions with the neutral 
 massive rho meson field $\rho_\mu$ and  
the  electromagnetic potential $A_\mu$ is described by the
  Lagrangian \cite{kapusta} 
\beq 
L 
= |D_\mu\Phi |^2
 - {1\over 4} \rho_{\mu\nu } \rho^{\mu\nu } + 
{1\over 2} m_\rho^2 \rho_{\nu }\rho^{\nu } 
- 
{1\over 4} F_{\mu\nu } F^{\mu\nu } 
\label{2.1}
\eeq
where $D_\mu\equiv \partial_\mu -ieA_\mu -ig_\rho\rho_\mu$ 
is the covariant derivative, $\Phi$ is the complex pion field, 
$\rho_{\mu\nu }$ is the rho and $F_{\mu\nu}$ is the 
photon field strength tensor. 
As is well known, the $\rho \pi \pi$ coupling is rather large
$g^2_\rho / 4 \pi \simeq 2.9$; nevertheless we attempt an
effective perturbative treatment up to two-loops.
However, no resummation of high temperature
effects, e.g. comparable to the hard thermal loop expansion
for QCD \cite{HTL,LeBellac}, is implemented in our approach,   
which maybe compared with an 
 analogous fixed-order calculation for  dilepton production
from a quark-gluon plasma \cite{baier}.

 From the Lagrangian (\ref{2.1}) and using the 
   closed-time-path formalism \cite{Lands,chou}
 we calculate  the dilepton production rate produced in a 
 thermal pionic medium at the two-loop level.
Typical diagrams are shown in Figs.~2 and 3.
The two tadpole diagrams of $O(g_\rho^2)$ are not included,
since they do not contribute to the discontinuity of the
photon selfenergy (\ref{2.3}).

\subsubsection{Real $\rho^0$ processes}

The processes involving real $\rho^0$'s, namely $\pi\pi\rightarrow \rho
\gamma^*$, $\pi\rho \rightarrow \pi\gamma^*$ and $\rho \rightarrow \pi\pi
\gamma^*$, are expected to be important for dilepton masses $M$ below the 
$\rho$-peak. These contributions are obtained by cutting the two-loop 
diagrams (Figs.~2 and 3) such that the $\rho$ is put on-mass shell. 

\begin{figure}[ht]
\centering
\epsfig{file=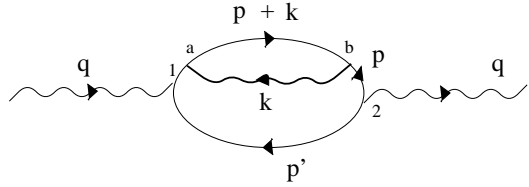, angle=-90, height=2.5 cm}
\caption{\label{figub}\it{Two-loop diagram: selfenergy insertion.
The labels a,b =1,2 denote the type of $\pi \rho$ vertex.
The solid line with momentum label k corresponds to the $\rho$}}
\end{figure}
%
%
%
\begin{figure}[ht]
\centering
\begin{minipage}[c]{10cm}
\centering
\centering
\epsfig{file=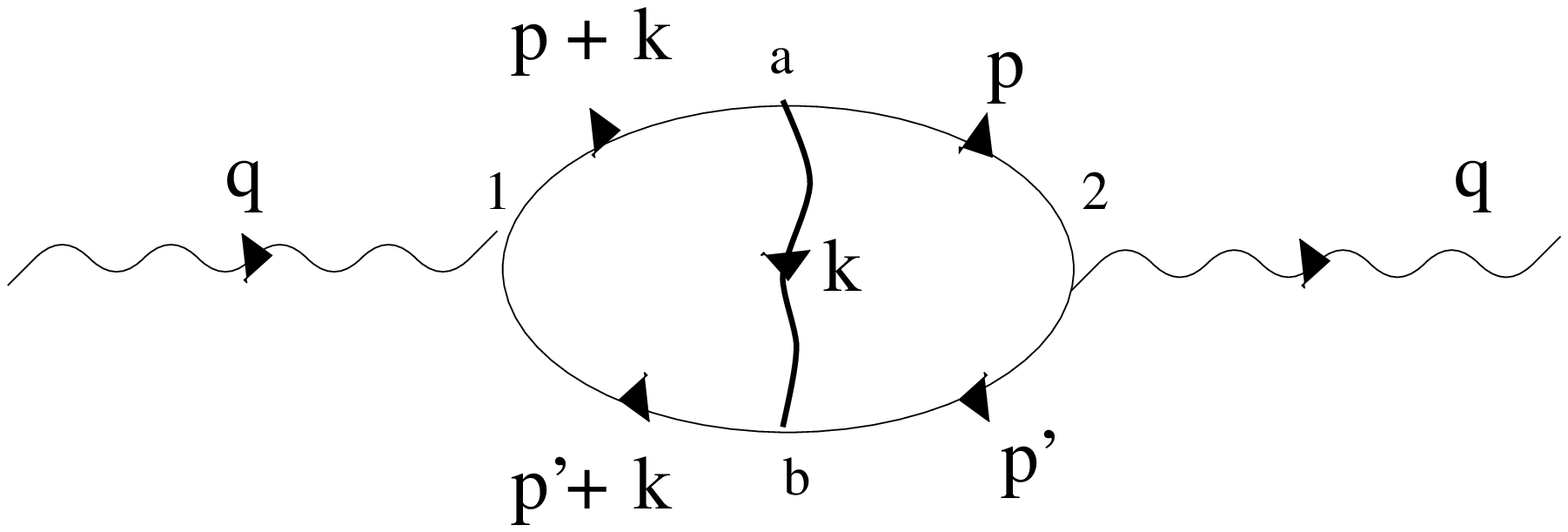, angle=-90, width=8 cm}
\end{minipage}%
\vskip .8cm
\begin{minipage}[b]{6.5cm} 
\centering
\centering
\epsfig{file=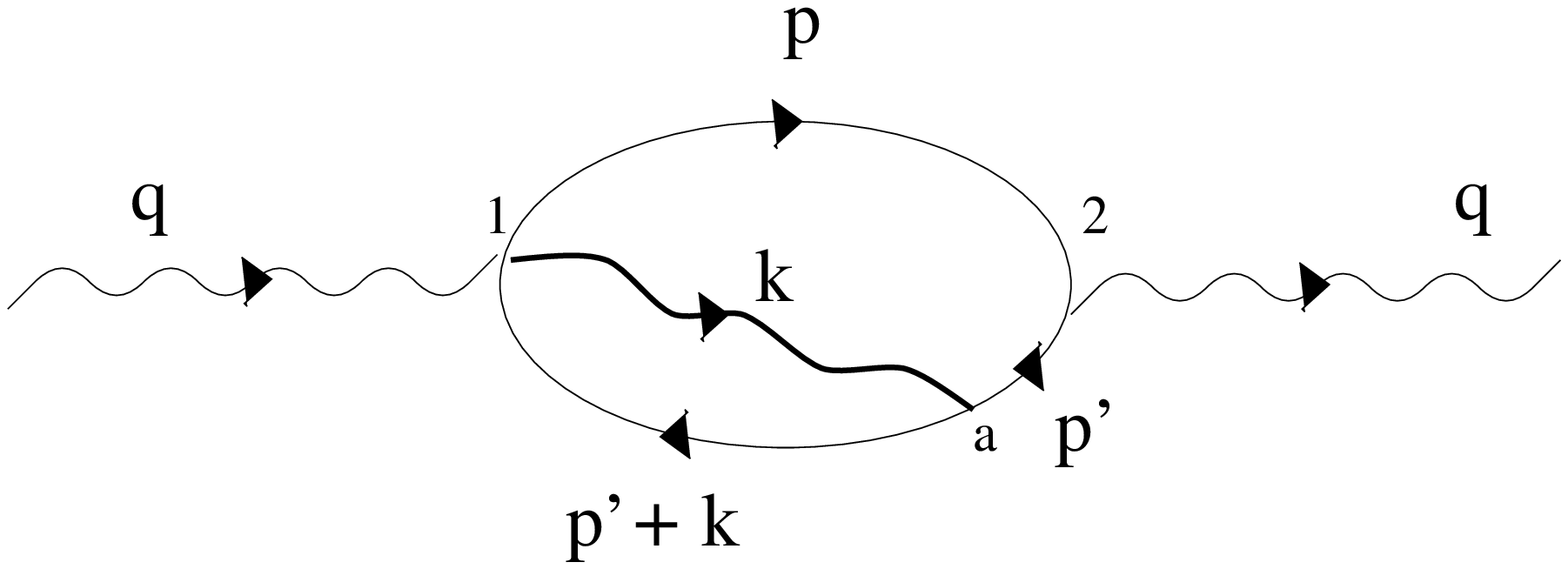, angle=-90, width =6cm}
\end{minipage}%
%
\begin{minipage}[b]{6.5cm}
\centering
\centering
\epsfig{file=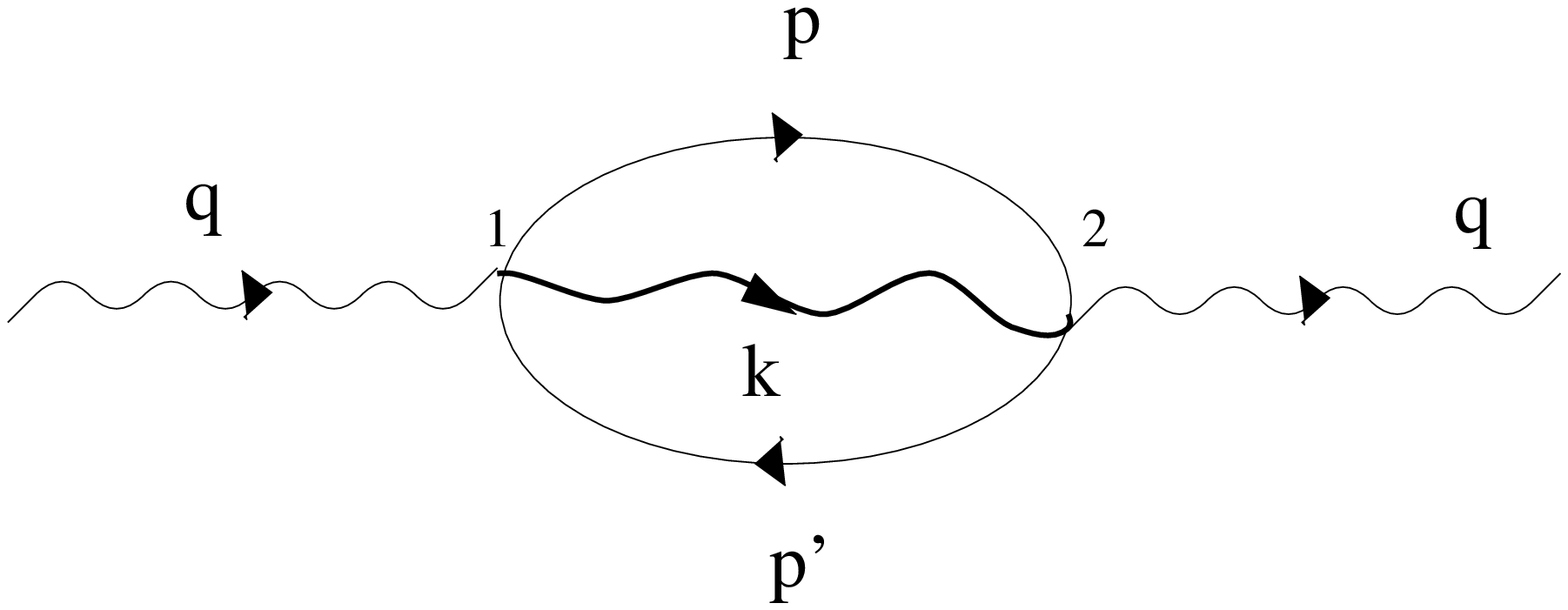, angle=-90, width=6cm}
\end{minipage}
\vskip .7cm
\caption{\label{fig4}\it{Two-loop diagrams: vertex type corrections}}
\end{figure}

We start to discuss explicitly the contribution from the diagram shown 
in Fig.~2, where the selfenergy correction $\delta D_{12} (p)$ is inserted into
the pionline with momentum $p$. The expression for $\delta D_{12} (p)$ is 
derived in Appendix B using the Dyson equation. 

In this section we assume (thermal and chemical) equilibrium distributions
(\ref{A.3}), and consequently we use $\delta D_{12}^{eq} (p)$ given in 
(\ref{B.5}). The discussion of out off equilibrium effects in the 
$\gamma^*$ rate is postponed to sect.~3.

 Using  the second term of (\ref{B.5}) we calculate the "real" correction
to (\ref{2.3a})
\beq
i \delta \Pi^{\gamma,SE}_{12} (q^0 , \vec q )  =   - (-ie)^2 \int \frac{d^4 p}
{(2\pi)^4} \, (p + p^\prime )^2 \,  
{\bf P} \left( \frac{1}{p^2}\right)^2  \, \left(-i \Pi_{12}
(p)\right) i D_{21} ( p^\prime ),
\label{2.9}
\eeq
where the pion selfenergy $i \Pi_{12} (p)$ at one-loop order due to 
$\pi\rho$ interactions (\ref{2.1}) is given by (cf. \ref{B.6} and \ref{C.2}), 
\beq
i\Pi_{12} (p) =  
 g^2_\rho \int \frac{d^4k}{(2\pi)^4} \, (2p+k)^\sigma \left(-g_
{\sigma\tau} + \frac{k_\sigma k_\tau}{m^2_\rho} \right) (2 p + k)^\tau 
\, D_{12} (p+k) D_{21} (k).
\label{2.10}
\eeq
We note that in (\ref{2.9}) we only keep the real $\rho^0$ contributions. 

Inserting the propagators (\ref{A.1}) with (\ref{A.9}) it is convenient to 
separate the different kinematical configurations, which appear due to the 
presence of the sign function in (\ref{A.9}). We consider in more detail
the process of $\rho$-emission, i.e. 
$\pi (p_+) + \pi (p_-) \rightarrow \rho (k) + \gamma^* (q)$, where we 
relabel to momenta of Fig.~2 as: $p^\prime \rightarrow -p_-,\, 
p\rightarrow p_+ -k$, and introduce (positive) energies $k^0 = E_\rho ,\,
p^0_- = E_- , \, p^0_+ = E_+$. 
We find the following contribution to $i\delta \Pi^{SE}_{12}$ of 
(\ref{2.9}), 
\beqar
& &\, + \,4 e^2 g^2_\rho \int \frac{d^3 p_+}{(2\pi)^3 2 E_+} \, n (E_+) \int 
\frac{d^3 p_-}{(2\pi)^3 2E_-} \, n (E_-) \int \frac{d^3 k}{(2\pi)^3 2 E_\rho}
 \,(1 + n (E_\rho))  \nn \\
& \times& (2\pi)^4 \delta^4 (p_+ + p_- - k - q) (2p_- - q)^2 \, {\bf P} 
\left( \frac{1}{(p_+ - k)^2} \right)^2  \,
 p^\sigma_+ \left( - g_{\sigma\tau} +
 \frac{k_\sigma k_\tau}{m_\rho^2} 
\right) p^\tau_+ ,  
\label{2.11}
\eeqar
expressed in a form familiar from kinetic theory \cite{mclerran}:
it consists of the square of the matrix element for the two-body process 
$\pi\pi \rightarrow \rho \gamma^*$, here to  one of
the exchange diagrams plotted in 
Fig.~4, with the integration over the phase-space of the participating 
pions and the $\rho$-meson properly weighted by the thermal distribution
functions $n (E)$. Next we also integrate with respect to the 
photon momentum $\vec q$ as described in Appendix D.

%
\begin{figure}[ht]
\centering
\epsfig{file=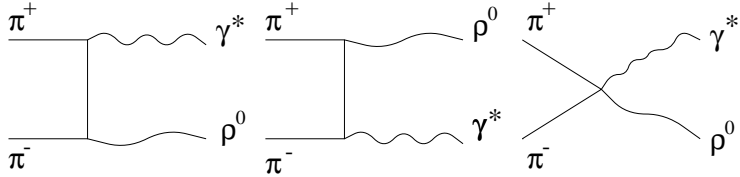, angle =-90, height=2.5cm}
\vskip .7cm
\caption{\label{matrix}\it{Matrix elements for
 the process  $~\pi \pi \to \rho
\gamma^*$ arising from cutting the two-loop diagrams}}
\end{figure}

An analogous, straightforward treatment gives the corresponding 
contributions to the real $\rho^0$ processes from the two-loop vertex type
diagrams of Fig.~3. Finally we sum all these contributions, including 
those obtained by interchanging momenta, like the selfenergy contribution
of (\ref{2.9}), but interchanging $p$ and $p^\prime$ in Fig.~2.

As a check the same final result
 is obtained when the square of the sum of the three
matrixelements illustrated in Fig.~4 is taken. This approach has e.g. 
the advantage that current conservation is rather easily checked on the 
level of these matrixelements.

Here we quote the result. Mainly for simplicity we use the 
Boltzmann approximation which amounts to underestimate the 
dilepton rate:
in this limit we replace e.g. in (\ref{2.11}) $n (E_+) n(E_-) \rightarrow
\exp ( -E_+ - E_-) /T$, 
and neglect stimulated emission, $1 + n (E_\rho) \simeq 1$. 
The dilepton rate due to the real $\rho^0$ processes is best expressed using 
the Mandelstam variables defined in (\ref{D.4}); together with the 
kinematical approximations discussed in Appendix D, we obtain for the 
dilepton mass range $M \leq m_\rho$, 
\beqar
 & &\frac{dN^{{\rm real}}}{dM^2 d^4 x}  \simeq  
\frac{\alpha^2 g^2_\rho / 4\pi}{24 \pi^4 M^2} \sqrt{\frac{\pi T^3}{2 m_\rho^3}}
\nn \\
&  &\times  \left\{ \left[ \int^\infty_{(m_\rho + M)^2} d s \, e^{-
\frac{s+m_\rho^2 - M^2}{2m_\rho T}} + \int^{(m_\rho - M)^2}_0 ds \, 
e^{-m_\rho / T} \right] \int^{u_+}_{u_-} du  \right. \,  
  + \left. 2\, {\bf P} \int^\infty_{m_\rho^2} dt \, e^{-\frac{t + m_\rho^2}
{2m_\rho T}} \int^{u_{{\rm max}}}_{u_{{\rm min}}} du \right\} \nn \\
&  & \times  \left[ 2 + \frac{m_\rho^2 M^2}{4} \left( \frac{1}{t^2} + 
\frac{1}{u^2} \right) + \frac{(m_\rho^2 + M^2)^2 + m_\rho^2 M^2/2}{tu} 
 -
 (m_\rho^2 + M^2) \left( \frac{1}{t} + \frac{1}{u} \right) \right], 
\label{2.12}
\eeqar
where 
\beqar
 & & u_{\buildrel +\over{(-)}} = {1 \over 2} \, (m_\rho^2 + M^2 -s)\, 
{\buildrel +\over{(-)}} \, {1 \over 2} \,
  \sqrt{(s-(m_\rho + M)^2) (s-(m_\rho - M)^2)} , 
 \nn \\
& &  u_{{\rm max}} = m_\rho^2 M^2 / t , \, ~~
 u_{{\rm min}} = m^2_\rho + M^2 - t , 
\label{2.12a}
\eeqar
with $ ~s+ t + u = m_\rho^2 + M^2$. 
The $u$-integration may still be performed analytically. 

The kinematic boundaries of integration given 
for the processes, which are summed in (\ref{2.12}) in the order of
$\pi \pi \to \rho \gamma^*, \rho \to \pi \pi \gamma^*,
\pi \rho  \to \pi \gamma^*$,  are illustrated in the 
Mandelstam plot of Fig.~8.  

The result 
contained in (\ref{2.12}),
 for the squared matrixelements of Fig.~4,
 is sucessfully compared with the one derived in \cite{kapusta}
for the case of the two-body processes involving a real photon.

It is important to note that due to the non-vanishing $\rho$ mass and the 
 heavy photon, $M>0$, no mass singularities appear in (\ref{2.12}),
even when setting the $T=0$ pion mass to zero. 
However, for the evaluation of the $\pi\rho \rightarrow \pi \gamma^*$ 
contribution, the principial value prescription leading to a well defined 
integral, e.g. in (\ref{2.12}) ${\bf P} ({1}/{u}) \equiv 
\lim_{\varepsilon\rightarrow 0} {u}/ ({u^2 +\varepsilon^2})$, and 
correspondingly for ${\bf P} (1/u^2)$ \cite{Lands}, has to be taken into 
account, in order to correctly treat the behaviour near $u \simeq 0$, which 
is covered by the kinematical domain of this process (Fig.~8). 
For the processes $\pi\pi \rightarrow \rho \gamma^*$ and $\rho \rightarrow
\pi\pi\gamma^*$ the ${\bf P}$ prescription is not explicitly required. 

\begin{figure}[ht]
\centering
\centering
\epsfig{bbllx=28mm,bburx=165mm,bblly=30mm,
bbury=155mm,file=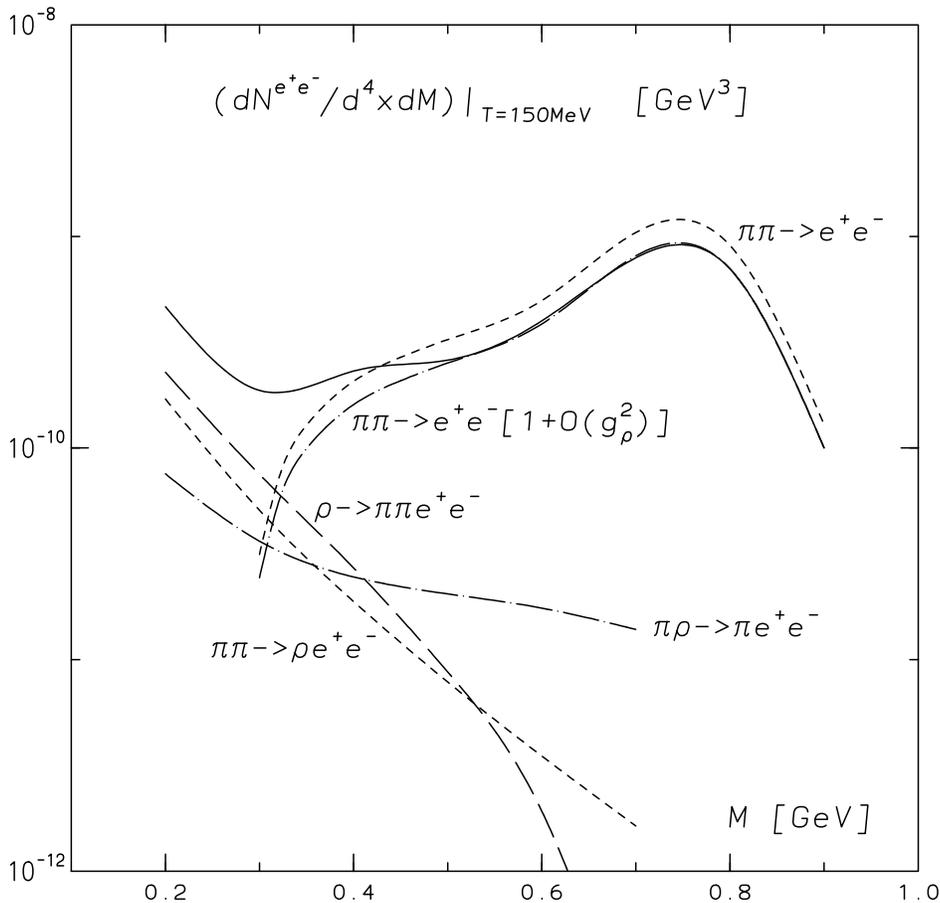, angle =-180}
\caption{\label{photona}\it{Dilepton rate as a function of $M$ at 
fixed $T = 150$ MeV. The solid curve represents the sum of  the real
and the virtual processes }}
\end{figure}
We plot the rate (\ref{2.12}) below the $\rho$-peak in Fig.~5 as a 
function of the dilepton mass $M$ for $M\geq 0.2$ GeV at fixed 
temperature $T=150$ MeV. For comparison the Born rate $\pi \pi \to 
\gamma^*$ is  also included. 
We find that, as expected, the two-body 
reactions $\pi\pi\rightarrow \rho\gamma^*$ and $\pi\rho\rightarrow \pi
\gamma^*$, together with the decay channel $\rho \rightarrow \pi\pi\gamma^*$, 
dominate the dilepton rate for low masses\footnote{We note that the curves
do not strongly change when the kinematical approximations performed in 
Appendix D are relaxed, e.g. for $\pi\pi\rightarrow \rho\gamma^*$ at most 
a 10\% change is numerically found when the $\rho$ is not kinematically 
kept fixed at rest}.
Thus, when discussing the thermal yield in the context
of recent experimental data   on dilepton production  in heavy ion collisions,
 it is necessary to include the above higher order processes.
 
 It is worth to mention that admitting in-medium effects on
 the $\rho$ meson mass or width, as recently proposed in Refs. [7,  9, 10],
 could substantially influence the overall thermal rate (solid curve)
 shown in Fig.~5. In particular, assuming a decrease  of the $\rho$ meson
 mass  by only $100$ MeV would imply an increase of the thermal rate
 (solid curve of Fig.~5) by a factor of 2.

\subsubsection{Virtual $\rho^0$ processes}

For heavy photon production at $O(g_{\rho}^2)$ we have,
in order to correct the Born rate (\ref{2.4}), to include
virtual contributions, which arise from the processes shown
in Figs.~2 and 3 by cutting the diagrams in the proper way
only through pion lines, without cutting the $\rho$ line.
 For the production of real
photons \cite{kapusta} these contributions are absent.

In some detail we describe our estimate, including the
approximations, for the selfenergy diagram (Fig.~2).
Here we take the first term of (\ref{B.5}) and evaluate
(cf. \ref{2.9}),
\beq
i \delta \Pi^{\gamma,virtSE}_{12} (q^0 , \vec q )  =   
 (-ie)^2 \int \frac{d^4 p^\prime}
{(2\pi)^3} \, (p + p^\prime )^2 \, \epsilon (p_0)~ n(p_0) 
\, \delta^{\prime}({p^2})  \, {\rm Re} \Pi(p) \, i D_{21} ( p^\prime ),
\label{2.20}
\eeq
where now the real part of the pion selfenergy $\Pi (p)$ enters,
and equilibrium distributions are used.

The temperature-independent part of ${\rm Re} \Pi$ is
 absorbed in the definition of 
the $T = 0$ pion mass (which we take approximately as $m_{\pi} = 0$).
The temperature dependent part in the one-loop case under 
consideration is  weighted by the thermal distribution either for the
$\rho$ meson or for the pion (see \cite{song}). The first case is
of $O(\exp(-m_\rho/T))$, i.e. negligible, the second one therefore
dominates and is expected to be of $O(T^2/m_{\rho}^2)$, due to the presence
of the $T = 0$ $\rho -$ propagator in the loop (Fig.~2).

In the following we estimate (\ref{2.20}) in the limit $m_\rho \gg T$,
having in mind the mass region $M \le m_\rho$.
To obtain the result analytically we make the further simplifying
assumption of dileptons at rest (cf. (\ref{2.6})). Then it is
straightforward to find
\begin{eqnarray}\label{2.21}
 && \hspace{4.5cm}
 i \delta \Pi^{\gamma,virtSE}_{12} (M , \vec q = 0 ) =  \nonumber \\
 &&  e^2   \int  \frac{d^3 p^\prime}{(2\pi)^2} 
 \frac{1}{2 E^\prime} ~ n(E^\prime)~ n(M - E^\prime) 
\; \delta^{\prime}(M/2 - E^\prime) ~(E^\prime / M - 1/4) 
 \,{\rm Re} \Pi(M - E^\prime, -\vec{p^\prime})~ , \hspace{2em}
 \end{eqnarray}  
where $| \vec{p^\prime} | = E^\prime$.

When evaluating ${\rm Re} \Pi (p_0, \vec{p})$ on-shell, we get
\beq
 {\rm Re} \Pi (M/2, \vec{p}) \simeq O(M^2 ~(T/m_\rho)^4),
\label{2.22}
\eeq
for $|\vec{p}| = M/2$, and therefore it will be  neglected.
 The leading term of
(\ref{2.21}) is found after partial integration with respect to the variable
$E^\prime$; it is proportional to the derivative 
\beq
{{\partial} \over {\partial E^\prime}}~   {\rm Re} \Pi 
 \simeq - {{ g_\rho^2}\over {4 \pi^2}}~
                                M~ \frac{T^2}{m_\rho^2}~,
\label{2.23}
\eeq
when evaluated on-shell because of the $\delta -$ function constraint
in (\ref{2.21}), and using Boltzmann distributions.
As expected we obtain 
\beq
i \delta \Pi^{\gamma,virtSE}_{12} (M , \vec q = 0 )  \simeq   
 - {{e^2} \over {8 \pi^2}}~  \frac{g_\rho^2}{4\pi}
~M^2~ {{T^2} \over {m_\rho^2}} \exp ( - M/T)~, 
\label{2.24}
\eeq
where we already include a factor 2 for the diagram Fig.~2
after interchanging $p$ with $p^\prime$.

In an analogous treatment we evaluate the virtual $T -$ dependent
contributions from the vertex type diagrams, namely from the 
first two of Fig.~3.
After a lengthy calculation the result is 
\begin{eqnarray}\label{2.cor}
 && \hspace{2.0cm} \frac{dN^{{\rm Born + virtual}}}{dM^2 d^4 x} 
  \simeq \nonumber \\ 
 &&  \frac{dN^{{\rm Born}}}{dM^2 d^4 x}  
\left[ ~ 1 + \frac{g_\rho^2}{4 \pi} ~(\frac{T}{m_\rho})^2 ~
( - \frac{19}{3 \pi} + \frac{M^2}{6 \pi T^2} ~
 {\bf P}  \int_0^\infty ~n(k)~ \frac{k dk}{(k^2 - M^2/4)})~ \right]
  \simeq \nonumber \\
 && \hspace{2.0cm}  \frac{dN^{{\rm Born}}}{dM^2 d^4 x}   
\left[~ 1 - \frac{7}{\pi}~ \frac{g_\rho^2}{4 \pi}
 ~(\frac{T}{m_\rho})^2 ~ \right]~,
 \end{eqnarray}  
valid for $m_\rho \gg T$, and for $M > 0$.

In Fig.~5 we see that e.g. for $T = 150$ MeV the  $T -$
dependent virtual corrections  (\ref{2.cor}) are 
negative and rather large. It 
suggests to perform resummations, which, however, we do not attempt.


\section{Out of equilibrium effects}\label{secoff}

In the early stage of heavy ion collisions, when lepton pairs are expected
to be predominantly produced \cite{shur,mclerran,ruuskanen,redl}, 
it is more likely that the pion gas is
not yet in thermal and chemical equilibrium \cite{Gavin,kaempfer}. 
Before we treat the 
realistic situation of an expanding gas in the next section, 
we first have to compute the production rate using non-equilibrium 
distributions, thus generalizing the calculation of sect.~2. 
We proceed with a tractable, but still realistic ansatz 
\cite{Gavin,kaempfer} by 
introducing the fugacity parameter \cite{groot}\ 
$\lambda$ and by using 
the following distributions\footnote{We do not distinguish the fugacity
parameter for $\pi^{\pm}$ and $\rho$ respectively}
$\tilde n(|k_0|) \simeq \lambda e^{-|k_0|/T}$ in Boltzmann approximation
(cf. (\ref{A.5}) and (\ref{A.6})). 
We have in mind that $\delta \lambda \equiv \lambda - 1$ 
is small $\delta \lambda \ll 1$ , i.e. a situation not far off 
equilibrium, in order to be consistent with the simplifying approximations 
summarized in Appendix A. 

In order to estimate non-equilibrium effects even under these approximations 
it is not justified to simply multiply the real emission rate 
(\ref{2.12}) 
by a factor $(1+2\delta\lambda )$ as it is in the case of the Born rate 
(cf. (\ref{2.4a})).
Due to the structure of the (one-loop) selfenergy correction to the pion 
propagator (\ref{B.3a}) a more careful derivation is required. 
For the real
emission contribution it first amounts to the replacement of (\ref{2.9}) 
by 
\begin{eqnarray}\label{3.1}
 i\delta \Pi_{12}^{\gamma,off}(q^0,\vec q)
 &\simeq& - (-ie)^2 \int\frac{d^4 p }
 {(2\pi)^4}\; (p+p')^2 \left\{ {\bf P}(\frac{1}{p^2})^2 \; n(p_0) \right. 
 \, (i \Pi_{12}(p) - i\Pi_{21}(p) ) \nonumber\\
 & &+ \left. \frac{1}{(p^2)^2 + (p_0\gamma)^2}
 \,(n(p_0) i \Pi_{21}(p) - (1+n(p_0))i \Pi_{12}(p) ) \right\} 
 i D_{21}(p'),
\end{eqnarray}  
where again 
a corresponding term with $(p \leftrightarrow p')$ has to be added.
In the second term in (\ref{3.1})
the pinch singularity, which is not cancelled in case of non-equilibrium 
distributions \cite{Altherr1}, is regularized \cite{Altherr2}
by the damping rate $\gamma$
of the pion, which we estimate in Appendix C. The expression for 
$i \Pi_{12}(p)$ is the same as in (\ref{2.10}), except that it now has 
to be 
evaluated using the non-equilibrium distributions (\ref{A.5}), 
as it is the 
case for $n(p_0)$ in (\ref{3.1}). 

We here discuss further details of the estimate for the dominant contribution, 
which is due to $\pi \rho \to \pi \gamma^\star$ in the limit $\gamma \to 
0$, in order to exhibit our treatment, i.e. the regularization
 of the pinch singularities
in the vincinity of $t/u \simeq 0$ (cf. Fig.~8). 

It is crucial to consider the following product of distributions, which is 
present in (\ref{3.1}), 
\begin{equation}\label{3.2}
 (1+n(p'_0)) \left[ n(p_0)n(k_0)(1+n(p_0+k_0)) - (1+n(p_0))(1+n(k_0))
 n(p_0+k_0) \right] ~,
\end{equation}
adjusted for the kinematics of the process $ \pi( p_{in})+ \rho ( k_\rho )
 \to \pi (p_{out}) +\gamma (q) $, i.e. $p'\to - p_{in}, k \to -k_\rho, 
 p \to - p_{out}+k_\rho $ (cf. Fig.~2).
 In relation to the definitions (\ref{D.4}) we 
have to continue: $ k \to - k_\rho , p_+ \to - p_{out}, p_- \to p_{in}$, 
which corresponds to the physical region (ii) with $ u \ge m_\rho^2 $
in Fig.~8. The limit $\gamma \to 0$  is sensitive to the 
behaviour near $ t= ( k_\rho - p_{out})^2 \simeq 0$. In the simplifying
kinematical approximation of the $\rho$ - meson at rest, $E_\rho \simeq
m_\rho$, used in Appendix D, the region $t \simeq 0$ corresponds 
to  positive energy $p_0 \simeq \frac{t+m_\rho^2}{2m_\rho} > 0$.

Thus the product of distributions  (\ref{3.2})
 reads in the Boltzmann approximation, 
\begin{equation}\label{3.3}
-\lambda^2
 (\lambda - 1) e^{\frac{E_{\rho}+E_{out}}{T}} \simeq - ~\delta \lambda
 e^{- \frac{u+m_\rho^2}{2m_\rho T}}
\end{equation}
valid for $p_0 > 0 $ and in leading order of  $\delta \lambda$ . 
The dominant contribution arising from (\ref{3.1}) to the 
dilepton spectrum becomes 
\begin{eqnarray}
& &\; \; \frac{dN^{pinch}}{dM^2 d^4 x } \stackrel{\gamma \to 0}
   {\simeq} -~ \delta \lambda 
 \,\frac{\alpha^2 (g_\rho^2/4\pi) }{48 \pi^4 M^2} \sqrt{\frac{\pi T^3}
 {2 m_\rho^3}}  \nonumber\\
 &  \times & \left\{ \int_{m_\rho^2}^{2m_\rho^2 + M^2}   du \;
  e^{- \frac{u+m_\rho^2}{ 2m_\rho T}}
  \int_{t_{min}}^{t_{max}}dt + \int_{2m_\rho^2 + M^2}^\infty du \;
 e^{-\frac{u+m_\rho^2}{2m_\rho T}}\int_{-m_\rho^2}^{t_{max}} dt \right\} 
 \;
 \frac{m_\rho^2M^2 }{t^2 + \left[ \frac{g_\rho^2T^2}{4\pi e} \right]^2 }~,
  \label{3.4}
\end{eqnarray}
where we take into account by a factor of 2 
both regions denoted by (ii) in Fig.~8.
The boundaries of the integrations are defined as in (\ref{2.12a}), 
except for 
$t \leftrightarrow u $. When using in (\ref{3.4}) the momentum 
independent 
value for the damping $\gamma$  (\ref{C.5}) 
the integrations can be explicitly
performed. The leading term with $\gamma \to 0$ is proportional to $1/
\gamma$, indicating the pinch singularity, 
\begin{equation}
 \frac{dN^{pinch}}{dM^2 d^4 x}\simeq 
- ~\delta \lambda \;
 \frac{\alpha^2 (g_\rho^2/4\pi)}
 {24 \pi^3} \sqrt{\frac{\pi T^3}{2m_{\rho}^3}} 
\; \frac{m_\rho^3 }{g_\rho^2/
 (4\pi e) T} \; e^{-\frac{2m_\rho^2 +M^2}{2m_\rho T}} ,
 \label{3.5} 
\end{equation}
which is actually independent of the coupling $g_\rho^2$.
 From Fig.~8 one can read off that the pinch near $t\simeq 0 $ 
is present starting at $ u \ge m_\rho^2 + M^2$. 
This and (\ref{3.3}) explains the argument of the 
exponential function in (\ref{3.5}). 

\begin{figure}[ht]
\centering
\centering
\epsfig{bbllx=28mm,bburx=165mm,bblly=30mm,
bbury=155mm,file=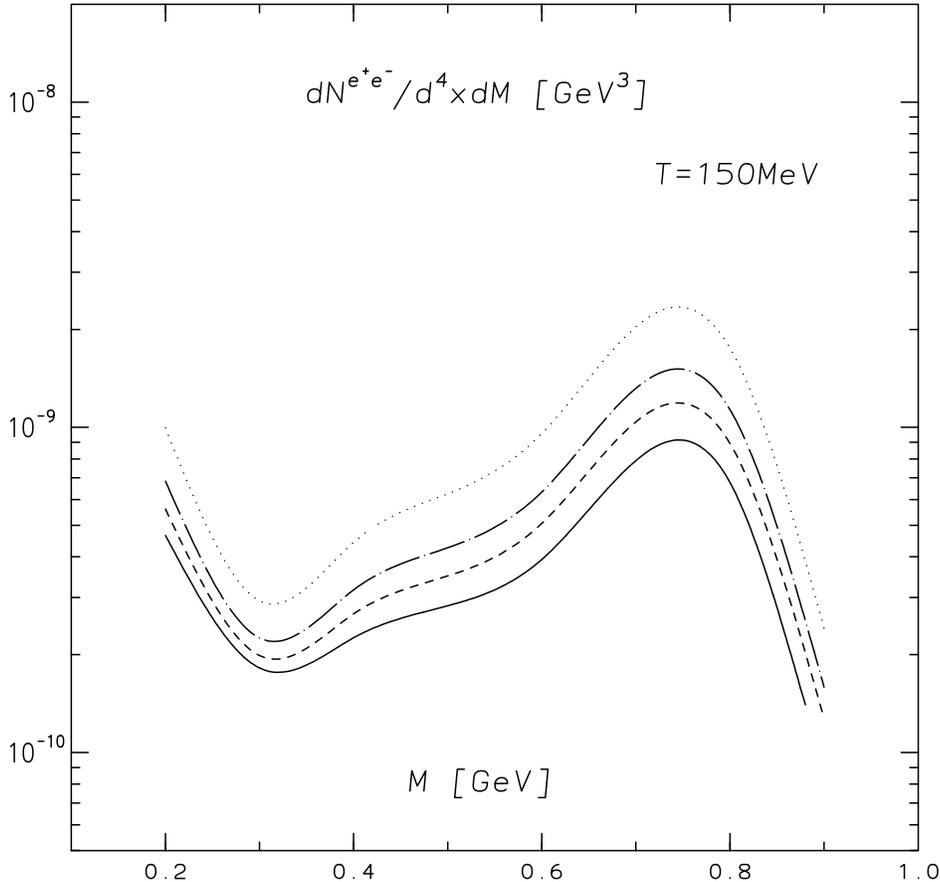, angle =-180}
\caption{\label{chemic}\it{Estimates of out of equilibrium effects to the
dilepton rate up to $O(g_\rho^2)$ 
at fixed $T = 150$ MeV:
solid curve corresponds to $\mu = 0$, dashed to $\mu = 25$,
dashed-dotted to $\mu = 50$, and dotted to $\mu = 100$ MeV,
respectively
}}
\end{figure}

We remark that (\ref{3.5}) differs numerically by only a 
few percent 
from the more exact expression starting from (\ref{3.1}) and including
all the contributions due to the processes (i) - (iii) defined in 
Appendix D. 

In Fig.~6 we plot the total rate $dN/dM d^4 x$, 
including the Born term 
with $O(g_\rho^2)$ correction, for non-equilibrium cases
 characterized by different 
values of the chemical potential $\mu$, including even as large values as
$\mu = 100$ MeV, which corresponds to a large value of $\lambda \simeq 1.95$
at $T=150$ MeV. Although here $\delta \lambda$ may be considered too large to 
justify keeping only $O(\delta\lambda)$ terms \protect\footnote{ 
A different estimate
in case of large $\lambda$, namely
 replacing $\gamma \to \lambda\gamma$, and
$Born(1+O(g_\rho^2)) \to \frac{\lambda^2 Born}{(1-O(\lambda g^2_\rho))}$, etc. 
actually gives the same result as in Fig.~6  }, 
as done in the calculation of the curves plotted in
Fig.~6, we observe that instead of an expected increase by a factor
of $\lambda^2 \simeq 4$, only an effective increase of the rate by a factor 2
results, because of the negative contribution estimated in (\ref{3.5}), 
when we change $\mu $ from $\mu =0$ to $\mu = 100$ MeV. From this fact we 
stress the importance of taking into account the non-trivial term 
(\ref{3.1}),
which is traced back to the selfenergy diagram (Fig.~2). 
For completeness we state that no such extra terms arise from the vertex type 
diagrams of Fig.~3. 


\section{Soft dileptons  in the expanding hadronic medium}

In the previous sections we have derived the dilepton production rate
originating from $\pi-\rho$ interactions up to $O(g_{\rho}^2)$ in a
    thermal medium as a function of temperature and for fixed
chemical potential.
In the description of dilepton spectra  in  heavy ion collisions
we have in addition to take into account that 
dilepton pairs are emitted from a rapidly expanding and thereby 
cooling medium.
Hot hadronic matter produced in heavy ion collisions
undergoes an
expansion, which leads to a space-time dependence of the thermal parameters.
In the expanding system we therefore have to integrate the
 rates derived in the previous sections  over
the space-time history of the collision. 
As a model for the expansion dynamics we assume
the Bjorken model for 1+1 dimensional longitudinal hydrodynamical
 expansion \cite{bj}.
In this model all thermodynamical parameters are only a function
of the proper time, $\tau$, and do not depend on the
 rapidity variable, $y$.
 We furthermore assume,
that at some initial time $\tau_0\sim 1$ fm      matter is
formed as a thermalized hadronic gas  with initial
temperature $T_0$.
The system subsequently expands and cools until it reaches
 the freeze-out temperature of $T_f\sim 130$ MeV.
We have modeled the hadronic  gas equation of state
with a resonance gas  where all hadrons and resonance states with
 masses up to $2.5$ GeV are included.
The value of the initial temperature $T_0\simeq 210$ MeV has been
fixed by the requirement to reproduce the multiplicity of 150
 charged pions in the final state at central rapidity.
These numbers are for S-Au collisions \cite{ceres}.
 
With the above defined model for the expansion dynamics
the space-time integration  of the rates,
first calculated in equilibrium, namely the Born rate
including the virtual correction of $O(g_\rho^2)$ (\ref{2.cor}), 
and the real processes (\ref{2.12}),
    can be performed leading to the spectrum which may
 than be compared with the one experimentally measured
in heavy ion collisions.
In Fig.~7 we show the overall thermal dilepton rate
from $\pi -\rho $ interactions including  acceptance
and kinematical cuts of the CERES experiment \cite{ceres}.
One  can see in Fig.~7 that with the  thermal
 source for dielectron pairs
due to $\pi-\rho$ interactions discussed in sect.~2
  one can partially account for
the access reported by the CERES Collaboration as measured 
 in S-Au (and Pb-Au)  collisions.
This is particularly the case in the vicinity of the
 $\rho$ peak where the agreement with 
the experimental data is quite  satisfactory.
However, the enhancement  below the $\rho$ peak  and the structure of the
 distribution  can not be explained by
equilibrium  production alone.
In the kinematical window $0.25<{\rm M}<0.45$ GeV there are still
almost $70\%$ deviations of our theoretical curve from the experimental
results.
Assuming non-chemical equilibrium in the
 mesonic medium (as discussed in sect.~3) should certainly
increase the dilepton rate. This is mostly because, in this case,
there are more  soft pions and rho mesons, which
should  produce more abundant soft   dileptons. In Fig.~7 we show the
thermal rate (long-dashed curve) assuming deviations from chemical
 equilibrium with the value of the chemical potential $\mu =100$ MeV.
Indeed, we observe an increase of the rate below the rho peak.
This increase, however, is
  rather modest particularly in the soft   part of the spectrum.
 The main reason  is due to the "pinch singular" term (\ref{3.5}),
 which being negative reduces the contributions  of the two-loop
 processes, when considered out of equilibrium.
In addition the out of equilibrium distributions of pions and rho mesons
lead to a lower initial temperature and a shorter lifetime of the
thermal system. All these effects are the reason of small
  modifications of  the  dilepton yield.

 Here, also additional assumptions e.g. on  in-medium effects 
 of the $\rho$ meson mass and width 
eventually help  to understand more completely
the dilepton access measured by the CERES Collaboration \cite{ko,gery,koch}.

\begin{figure}[ht]
\centering
\centering
\epsfig{bbllx=28mm,bburx=165mm,bblly=30mm,
bbury=155mm,file=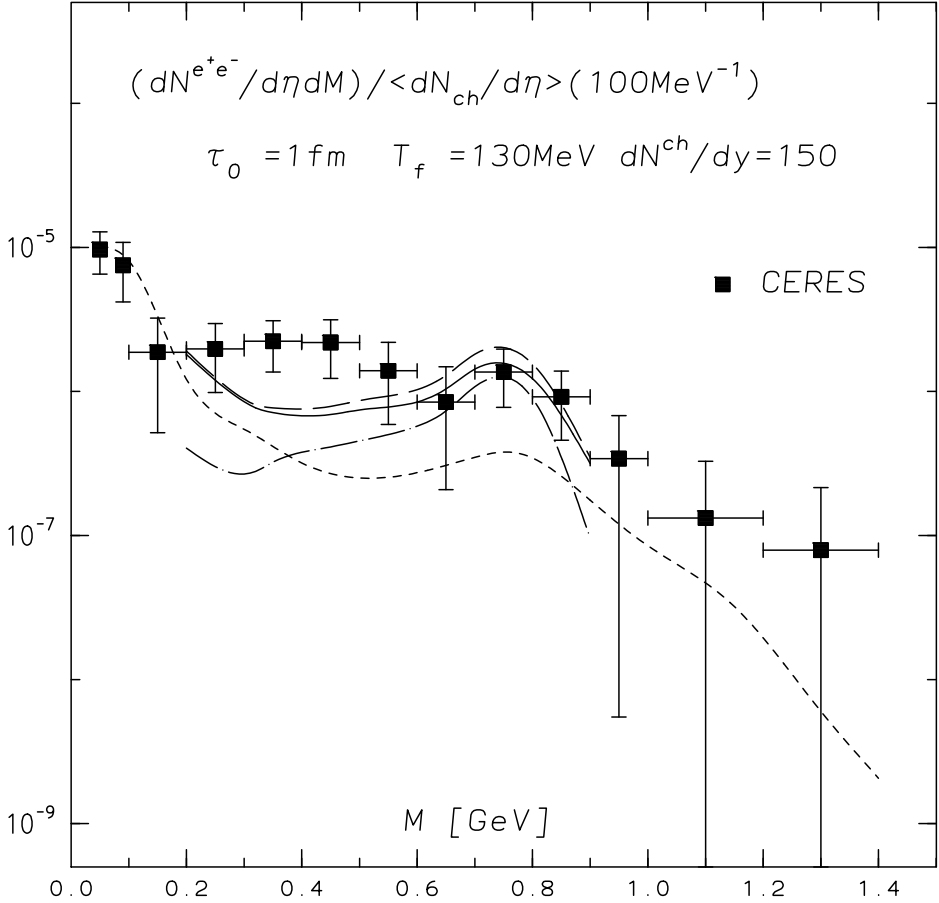, angle =-180}
\caption{\label{final}\it{
Invariant-mass  dielectron spectra measured by the CERES Collaboration
 for the
S-Au collisions \cite{ceres} in comparison with the thermal yield from
$\pi -\rho$ interactions.   Dashed-dotted line describes the sum
of all contributing reactions  from $\pi^\pm - \rho^0$ interactions.
The dashed line, calculated by CERES,  represents the expected dielectron
yield from all known hadronic sources. The  full-line is the sum
of the contributions described by the
 dashed-dotted and dashed curves.
The long-dashed curve includes the out of equilibrium effects
}}
\end{figure}


\section{Summary }

We have calculated the thermal production rate for soft 
dielectrons produced in a pion gas including  relevant 
 reactions from $\pi-\rho$ interactions.
 The calculations include the  $O(g_{\rho}^2)$ corrections 
 arising from the two-loop contributions to the virtual photon 
selfenergy. The dilepton rate is first calculated
using equilibrium distributions. Next off equilibrium
contributions to the rates have been estimated.
Finally, dilepton production in an expanding hadronic medium
has been calculated  applying the Bjorken model.
No medium effects on he $\rho$ resonance parameters are
included in the analysis.
The access reported by the CERES Collaboration in the
low mass range below the $\rho$ peak is diminished,
although not yet completely described by the model.

\vspace{0.5cm} 
\noindent 
\subsection*{Acknowledgements} 
 
One of us K.~R. acknowledges stimulating discussion with members of the
CERES Collaboration and partial support
of the Gesellschaft f\"ur Schwerionenforschung (GSI) and 
of the Committee of Research Development 
(KBN 2-P03B-09908). 
Supported in part by the EEC Programme "Human Capital and Mobility",
Network "Physics at High Energy Colliders", Contract CHRX-CT93-0357.
M.~D. is supported by DFG.
\newpage


\begin{appendix}

\renewcommand{\theequation}{\thesection.\arabic{equation}}
\makeatletter
\@addtoreset{equation}{section}
\makeatother

\setcounter{equation}{0}
\renewcommand{\theequation}{A.\arabic{equation}}

\subsection*{Appendix A: Thermal propagators}\label{AppA}

In order to perform the perturbative calculations presented in this paper 
we use the closed-time-path formalism of thermal field theory 
\cite{Lands,chou,Eijck}.
The real-time $2\times 2$ matrix (scalar) propagators,   
\beqar
D_{11} (k) & = & (1+n(k_0)) \Delta_R (k) - n (k_0) \Delta_A (k) , \nn  \\
D_{12} (k) & = & n (k_0) (\Delta_R (k) - \Delta_A (k)), \nn \\
D_{21} (k) & = & (1 + n (k_0)) ( \Delta_R (k) - \Delta_A (k)), \nn \\
D_{22} (k) & = & n (k_0) \Delta_R (k) - (1+n (k_0)) \Delta_A (k), 
\label{A.1}
\eeqar
are expressed in terms of the retarded and advanced propagators, 
\beq
\Delta_{R (A)} (k) = \frac{1}{k^2 - m^2 {+\atop {(-)}} i \epsilon k_0}, 
\label{A.2}
\eeq
where $k^2 = k^2_0 -{\vec k}^2$ and $m$ is the 
$T = 0$ mass.
For equilibrium conditions the 
Bose-Einstein distribution function $n (k_0)$ depends on 
temperature $T$, 
\beq
n(k_0) = \frac{1}{e^{k_0 / T} - 1} , 
\label {A.3}
\eeq
and satisfies 
\beq
n (k_0) + n (-k_0) + 1 = 0 . 
\label{A.4}
\eeq
For our numerical estimates we are mainly using the Boltzmann
approximation
\beq
n (|k_0|) \simeq \exp (- |k_0|/T) . 
\label{A.4a}
\eeq

Estimating non-equilibrium effects we follow the approximations described
in detail in \cite{chou}, and more recently in \cite{LeBellac2}.
It amounts to  replace the thermal $n(k_0)$, (\ref{A.3}), by
their non-equilibrium counter parts, i.e. in general by Wigner 
distributions $n (k_0, X)$.
This corresponds in terms of the cumulant expansion to approximate 
non-equilibrium correlations by the second cumulant only. As a further
approximation we suppress the possible dependence on the center of mass 
coordinate $X$, essentially assuming a homogenous and isotropic medium.

Practically for our estimates we take (cf. (\ref{A.4})), 
\beq
n(k_0,X) \rightarrow 
\left\{  \begin{array}{l@{\quad ~ \quad}l} 
\tilde n (|k_0|)     & k_0 > 0 , \\
- (1 + \tilde n (|k_0|) & k_0 < 0 ,
\end{array} \right.
\label{A.5}
\eeq
with
\beq 
\tilde n (|k_0|) \equiv \frac{1}{e^{(|k_0| -\mu)/T} -1} \hat= \frac{\lambda}
{e^{|k_0|/T} -\lambda} , 
\label{A.6}
\eeq
by introducing the fugacity parameter \cite{groot} 
$\lambda \equiv e^{\mu/T}$, which is
assumed to be energy independent. Obviously $\lambda \not= 1$ in case of
(chemical)  non-equilibrium. Consequently deviations from equilibrium are 
approximated in terms of the equilibrium distribution (\ref{A.3}) by
\beq
\frac{\delta n}{n(k_0)}  \equiv  n (k_0, X)/ n (k_0) -1 \simeq 
  {\delta\lambda}  \, \epsilon (k_0) \, (1 + n (k_0)) 
\label{A.8}
\eeq
in leading order of $\delta\lambda \equiv \lambda -1$.

In the expressions for $\Delta_{R(A)} (k)$ the usual limit
$\epsilon \rightarrow 0$ is considered, whenever this limit is defined,
e.g.
\beq
\Delta_R (k) - \Delta_A (k) = - 2 \pi i \epsilon (k_0)
 \delta (k^2 - m^2) ,
\label{A.9}
\eeq
 where $\epsilon
(k_0) = \theta (k_0) - \theta (-k_0)$ is the sign function. 

However, out of equilibrium pinch singularities appear in this limit 
$\epsilon \rightarrow 0$ \cite{Altherr1}, e.g. in products like $\Delta_R (k)
\Delta_A (k)$ (see Appendix B). We treat this situation by following the 
 conjecture by Altherr \cite{Altherr2,Bedaque} keeping 
$\epsilon$ non-vanishing: it is identified by the damping width 
$\gamma > 0$ (Appendix C).


\setcounter{equation}{0}
\renewcommand{\theequation}{B.\arabic{equation}}
\subsection*{Appendix B: One-loop selfenergy correction to the pion propagator}

Here we comment (cf. Fig.~2) on the selfenergy correction of the pion
propagator and the possible appearence and presence of pinch  
singularities \cite{Lands,Altherr1}. Denoting the selfenergy 
insertion by the matrix $\Pi_{ab} (p)$ we obtain the improved 
$(12)$-propagator using the Dyson equation,
\beqar
i D_{12} (p) &\rightarrow & i D_{12} (p) + \sum_{a,b =1,2}
 i D_{1a} (p) (-i
\Pi_{ab}(p)) i D_{b2} (p) \nn \\
& = & i D_{12} (p) + i \delta D_{12} (p). 
\label{B.1}
\eeqar
With the propagators specified in the Appendix A, and with the relations 
\beq
\Pi_{11} (p) = - \Pi^\star_{22} (p), ~~ {\rm Im}\, \Pi_{11} = \frac{i}{2} 
(\Pi_{12} + \Pi_{21}),
\label{B.2}
\eeq 
also valid out of equilibrium, we obtain
\beqar
\delta D_{12} (p) &=& n (p_0) (\Delta^2_R (p) - \Delta^2_A (p))
{\rm Re} \, \Pi_{11} (p) \nn\\
&+& \frac{1}{2} n (p_0) (\Delta^2_R (p) + \Delta^2_A (p)) 
[ \Pi_{12}(p) - \Pi_{21}(p)] \nn \\
&+& \Delta_R (p) \Delta_A (p)
 [ n(p_0) \Pi_{21}(p) - (1 + n (p_0)) \Pi_{12}(p) ].
\label{B.3a}
\eeqar
Here the ill-defined product $\Delta_R (p) \Delta_A (p)$ appears, giving 
rise to possible unpleasant pinch singularities \cite{Lands,Altherr1}. 
In case of equilibrium, however, it is well known that the last term in 
(B.3) 
vanishes due to the detailed balance condition, 
\beq
\Pi_{21} (p) = e^{p_0/T} \Pi_{12} (p). 
\label{B.4}
\eeq
As a consequence
(B.3) simplifies to 
\beq
\delta D_{12}^{eq} (p) =  2\pi i \epsilon (p_0) n (p_0)\left[ \delta^\prime 
(p^2) {\rm Re}\, \Pi (p)  
+ \frac{1}{\pi} {\bf P} \left(\frac{1}{p^2}\right)^2 {\rm Im}\,\Pi (p) \right],
\label{B.5}
\eeq
where
\beq
{\rm Re}\, \Pi (p) \equiv {\rm Re}\, \Pi_{11} (p), \, ~~\,
 {\rm Im}\,\Pi \equiv 
\frac{1}{2} \epsilon (p_0) \frac{i\Pi_{12} (p)}{n(p_0)} ,
\label{B.6}
\eeq
 expressed in terms of the equilibrium distribution $n(p_0)$,
and in terms of the (retarded) selfenergy 
$\Pi(p_0 + i\epsilon p_0, \vec p)$ \cite{LeBellac}.
$\delta^\prime$ denotes the derivative of the $\delta$-function, and ${\bf P}$
the principal value. 

However, for nonequilibrium distributions (\ref{A.5}), for which obviously
$(1+n)/n \not= \exp (p_0/T)$, the last term in (B.3)
 does not cancel!
In order to regularize the product $\Delta_R (p) \Delta_A (p)$
\cite{Altherr2} we take into account the non-vanishing (on-shell) damping
rate of the pion, which we estimate in the following Appendix C.


\setcounter{equation}{0}
\renewcommand{\theequation}{C.\arabic{equation}}
\subsection*{Appendix C: Estimate of the damping rate of the pion}\label{AppC}

The pion damping rate $\gamma$ is determined by the ``pole'' (in the lower 
energy half-plane) of the retarded propagator. In the following
we not only neglect the pion mass, but also thermal corrections due to the 
real part of the (retarded) pion selfenergy, such that 
\beq
\gamma \simeq - \frac{{\rm Im}\, \Pi (p_0, 
{\vec p})}{p_0} , 
\label{C.1}
\eeq
evaluated on-shell $p^2 = 0$ for positive pion energy $p_0$. In the 
one-loop approximation the absorptive part of the 
pion selfenergy is given by
\beq
{\rm Im}\,\Pi (p_0 ,{\vec p}) = 
\frac{g^2_\rho}{2n (|p_0|)} \, \int \frac{d^4 k}{(2\pi)^4} ( 2 p + k)^\sigma
\left( -g_{\sigma\tau} + \frac{k_\sigma k_\tau}{m^2_\rho} \right)
 (2 p + k)^\tau \,
 D_{12} (p + k) D_{21} (k), 
\label{C.2}
\eeq
where the momentum labels of Fig.~2 are used. 

The temperature dependent non-vanishing part for $p^2 = p^2_0 - 
{\vec p}^2 = 0$, becomes (cf. the derivation in \cite{song}),
\beq
{\rm Im}\,\Pi (p_0 = p, | {\vec p} | 
= p) = 
-g^2_\rho m^2_\rho \pi \int \frac{d^3 k}{(2\pi )^3} 
\frac{1}{2 E_p 2 E_\pi} \,
 [n (E_\pi ) - n ( E_\rho)]\, \delta (p-E_\rho + E_\pi), 
\label{C.3}
\eeq
with the corresponding energies 
are $E_\rho = \sqrt{{\vec k}
^2 + m^2_\rho}$ and $E_\pi = | {\vec p} + {\vec k}| $. 

The dominant contribution comes from the pion's thermal distribution 
$n(E_\pi)$. In the  Boltzmann approximation it leads to
\beq
{\rm Im}\, \Pi (p_0 = p , p) \simeq - \frac{g^2_\rho}{4\pi}
\frac{m^2_\rho}{4p} \int^\infty_{\frac{m^2_\rho}{4p}} n (E_\pi) d E_\pi
\simeq - 
\frac{g^2_\rho}{4\pi} \frac{m^2_\rho T}{4p} e^{-\frac{m^2_\rho}{4pT}} .
\label{C.4}
\eeq
${\rm Im}\, \Pi$ vanishes for ${\vec p} = 0$, and it has its
maximum near $p\simeq \frac{m^2_\rho}{4T}$. We note that (\ref{C.4}) gives
a positive damping rate $\gamma$, as required. In order not to 
overestimate the contributions arising from the $\Delta_R\Delta_A$ terms we 
take for the numerical estimates a momentum independent value,
namely the one at the maximum,
\beq 
p_0 \gamma \simeq - {\rm Im}\, \Pi\,(p_0 \simeq p, p) \simeq 
\frac{g^2_\rho}{4\pi}\frac{1}{e} T^2 .
\label{C.5}
\eeq
 

\setcounter{equation}{0}
\renewcommand{\theequation}{D.\arabic{equation}}
\subsection*{Appendix D: Kinematic boundaries for the kinetic processes}
\label{AppD}

The dilepton thermal rate receives contributions from the two-body processes
$\pi\pi \rightarrow \rho \gamma^*$ and $\pi\rho \rightarrow \pi \gamma^*$, 
and from $\rho \rightarrow \pi \pi \gamma^*$, for which we evaluate in the 
following the phase-space integral (\ref{2.11}) including its kinematic 
constraints. The calculation is simplified using 
(i) Boltzmann distributions and 
(ii) the fact that $m_\rho /T \gg 1$: the $\rho$-meson is kept at rest in the 
medium. This allows to integrate the $\rho$'s phase-space by
\beq
I_\rho \hat= \int \frac{d^3k}{(2\pi)^3} \frac{1}{2E_\rho} e^{-E_{\rho}/T}
\simeq \frac{T^2}{4\pi^2} \sqrt{\frac{\pi m_\rho}{2T}} e^{-m_{\rho}/T},
\label{D.1}
\eeq
and to replace elsewhere $E_\rho = \sqrt{{\vec k}^2
+m^2_\rho} \simeq m_\rho$.

The remaining integrations are carried out as we explicitly show 
in the following for the 
channel $\pi (p_+) + \pi (p_-) \rightarrow \rho (k) + \gamma^* (q)$.
We only consider the case of dilepton masses $M\leq m_\rho$. 

The phase-space integral  defined by 
\beq
I  \hat=  \int \frac{d^3 q}{q_0}
 \int \frac{d^3 p_+}{(2\pi)^3 2E_+} \int \frac
{d^3p_-}{(2\pi)^3 2E_-} \int \frac{d^3 k}{(2\pi)^3  2 E_\rho} e^{-E_{\rho}/T}
 \,
(2 \pi)^4 \, \delta^4 \, (p_+ + p_- - k - q)
\label{D.2}
\eeq
is approximated by
\beq
I \simeq  {1 \over {4 \pi^2}} I_\rho
 \int \frac{d^3 p_+}{ 2 E_+} \int \frac
{d^3p_-}{ 2 E_-} {1 \over q_0}
\delta \, (E_+ + E_- - m_\rho - q_0)
\label{D.3}
\eeq
with $q_0 \simeq \sqrt{ |{\vec p}_+ + 
{\vec p}_- |^2 + M^2}$. 
The angular integrations in $I$
can be done.

\begin{figure}[ht]
\centering
\epsfig{file=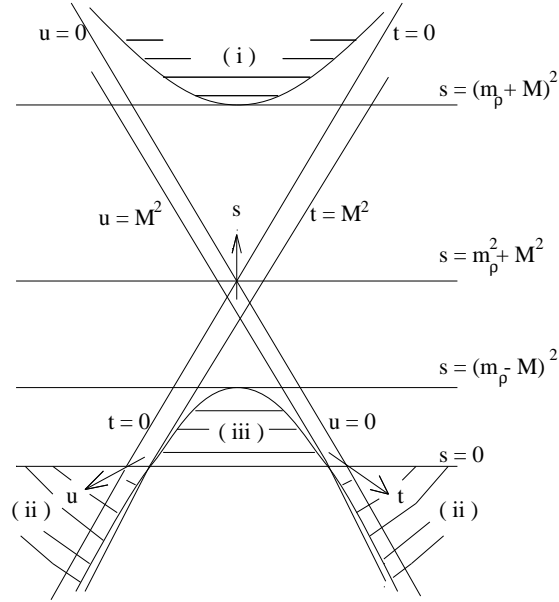, angle =-90, height=8cm}
\vskip .7cm
\caption{\label{mandel}\it{Mandelstam plot:
physical regions for the processes (i) $~\pi \pi \to \rho
\gamma^*$, (ii) $~ \pi \rho \to \pi \gamma^*$ and (iii)
$~\rho \to \pi \pi \gamma^*$}}
\end{figure}

 Next 
we introduce the Mandelstam variables by
\beqar
 s = (p_+ + p_-)^2 & = & (k+q)^2 , \nn \\
 t = (p_+ - k)^2 & = & (q-p_-)^2 , \nn \\
 u = (p_- -k)^2  & = & (q-p_+)^2 , 
\label{D.4}
\eeqar
with $s + t + u = m_\rho^2  + M^2$.
Approximating the pion energies by 
\beq
E_+ \simeq (m_\rho^2 -t) /2m_\rho, \,\,\, E_- \simeq (m_\rho^2 - u)/ 2m_\rho , 
\label{D.5}
\eeq 
gives the final result for the phase-space integral,
\beq
I\simeq \frac{1}{8 m_\rho^2} \, I_\rho  \int dt du  \, \theta
(t u - m^2_\rho M^2) \, \theta (s - (m_\rho + M)^2).
\label{D.6}
\eeq
The other channels only differ in their kinematical boundaries,
\beqar
(ii)\,\, \pi\rho \rightarrow \pi\gamma^* &:& \theta (-s)\, \theta
  (m^2_\rho M^2 - tu), \nn \\
(iii)\,\, \rho \rightarrow \pi\pi\gamma^* & : & \theta (s) \, \theta (tu - m^2_
\rho M^2), 
\label{D.7}
\eeqar
 which are 
also summarized in Fig.~8.
The processes $\pi \pi \rho \to \gamma^*$ and $\pi \to \pi \rho \gamma^*$
do not contribute due to the nonvanishing masses $m_\rho$ and $M$.

We finally
remark that the poles at $t= 0$ and/or $u=0$ are present within the 
kinematical domain for the channel $\pi\rho \rightarrow \pi\gamma^*$, whereas
for $\pi\pi \rightarrow \rho\gamma^*$  
they are present just at the kinematical boundary when $s \to \infty$. 

\end{appendix}


\newpage
 
\begin{thebibliography}{99}
\bibitem{shur}
 E.~L.~Feinberg, Nuovo Cimento {\bf 34A} (1976) 391;\\ 
 E.~Shuryak, Phys. Lett. {\bf 78B} (1978) 150.
%
\bibitem{mclerran}
 L.~McLerran and T.~Toimela, 
 Phys. Rev. {\bf D31} (1985) 545. 
\bibitem{ruuskanen}
P.~V.~Ruuskanen, in {\it 
 Quark-Gluon Plasma},  ed. R.~Hwa (World Scientific, Singapore, 1990);\\
J.~Alam, S.~Raha and B.~Sinha, Phys. Rep. {\bf 273} (1996) 243.
%
\bibitem{redl}
 J.~Cleymans, K.~Redlich and H.~Satz, Z. Phys. {\bf C52} (1991) 517. 
%
\bibitem{ceres}
 CERES Collab., G.~Agakichiev et al., 
 Phys. Rev. Lett. 
 {\bf 75} (1995) 1272; \\
P.~Holl et al., preprint CERN-SPSLC-95-35 (June 1996).
\bibitem{helios}
 HELIOS/3 Collab., M.~Masera, 
Nucl. Phys. {\bf A590} 
(1995) 3c. 
\bibitem{ko}
  G.~Q.~Li, C.~M.~Ko and G.~E.~Brown, Phys. Rev. Lett. {\bf 75} (1995) 4007;\\ 
%
W.~Cassing, W.~Ehehalt, C.~M.~Ko, Phys. Lett. {\bf B363} (1995) 35.
\bibitem{dinesh} 
D.~K.~Srivastava, B.~Sinha and C.~Gale,  
Phys. Rev.  {\bf C53} (1996) R567.
\bibitem{gery}
%
 C.~Adami and G.~E.~Brown, Phys. Rep. 
 {\bf 224} (1993) 1;\\
T.~Hatsuda and S.~H.~Lee, Phys. Rev. {\bf C46} (1992) R34; \\
G.~Q.~Li and C.~M.~Ko, 
Nucl. Phys. {\bf A582} (1995) 731; \\
F.~Karsch, K.~Redlich and L.~Turko, Z. Phys. {\bf C60} (1993) 519;\\ 
R.~D.~Pisarski, Phys. Rev. {\bf D52} (1995) R3773. 
\bibitem{koch}
C.~Song, V.~Koch, S.~H.~Lee and C.~M.~Ko,
 Phys. Lett. {\bf B366} (1996) 379. 
\bibitem{gale}
 C.~Gale and J.~I.~Kapusta, 
 Nucl. Phys. {\bf B357} (1991) 65;  Phys. Rev. {\bf C35} (1987) 2107. 
\bibitem{gale2}
 C.~Gale and P.~Lichard, 
   Phys. Rev. {\bf D49} (1994) 3338. 
%
\bibitem{song} 
C.~Song, Phys. Rev. {\bf D49} 
 (1994)     1556; Phys. Rev. {\bf D48} (1993)        1375. 
\bibitem{kapusta}
 J.~I.~Kapusta, P.~Lichard and 
 D.~Seibert, Phys. Rev. 
{\bf D44} (1991) 2774. 
\bibitem{pisarski}
  E.~Braaten, R.~D.~Pisarski and T.~C.~Yuan, 
Phys. Rev. Lett. {\bf 64} (1990) 2242;\\
S.~M.~H.~Wong, Z. Phys. {\bf C53} (1992) 465. 
%
\bibitem{weldon}
 H.~A.~Weldon, Phys.Rev. 
 {\bf D42} (1990) 2384. 
\bibitem{weldon1}
 H.~A.~Weldon,  Phys. Rev. {\bf D28} (1983) 2007;\\
%
 R.~L.~Kobes and S.~W.~Semenoff, 
 Nucl. Phys. {\bf B260} (1985) 714. 
%
\bibitem{Lands}
N.~P.~Landsman and Ch.~G.~van Weert, Phys. Rep. {\bf 145} (1987) 141.
\bibitem{chou}
K.-C.~Chou, Z.-B.~Su, B.-L.~Hao and L.~Yu, Phys. Rep. {\bf 118} (1985) 1.
\bibitem{Eijck}
M.~A.~van~Eijck and Ch.~G.~van~Weert, Phys. Lett. {\bf B278} (1992) 305.
\bibitem{Haglin}
K.~L.~Haglin, Phys. Rev. {\bf C53} (1996) R2606; \\
see also: C.~M.~Hung and E.~V.~Shuryak, preprint SUNY-NTG-96-16,
hep-ph/9608299.
\bibitem{HTL}
R.~D.~Pisarski,
Phys. Rev. Lett. {\bf 63} (1989) 1129;\\
E.~Braaten and R.~D.~Pisarski, Nucl. Phys. {\bf B337} (1990) 569;\\
R.~D.~Pisarski,
 Nucl. Phys. {\bf A525} (1991) 175c, 
and references  therein. 
%
\bibitem{LeBellac}
M.~Le~Bellac, {\it ``Thermal Field Theory''}, (Cambridge University Press,
Cambridge, 1996), and references therein.
\bibitem{baier} 
See, e.g. : R.~Baier, B.~Pire and D.~Schiff, 
 Phys. Rev. {\bf D38} (1988) 2814; \\
T.~Altherr and P.~Aurenche, Z. Phys. {\bf C45} (1989) 99.  
\bibitem{Gavin}
S.~Gavin, Nucl. Phys. {\bf B351} (1991) 561; \\
S.~Gavin and P.~V.~Ruuskanen, Phys. Lett. {\bf B262} (1991) 326. 
\bibitem{kaempfer}
P.~Koch, Z. Phys. {\bf C57} (1993) 283; \\
B.~K\"ampfer, P.~Koch and O.~P.~Pavlenko, Phys. Rev. {\bf C49}
(1994) 1132.
\bibitem{groot}
S.~R.~de Groot, W.~A.~ van Leeuwen and Ch.~G.~van Weert,
{\it ``Relativistic Kinetic Theory''}, (North Holland,
Amsterdam, 1980); \\
see also: T.~S.~Biro et al.,  Phys. Rev. {\bf C48} (1993) 1275.
%
\bibitem{Altherr1}
T.~Altherr and D.~Seibert, Phys. Lett. {\bf B333} (1994) 149.
\bibitem{Altherr2}
T.~Altherr, Phys. Lett. {\bf B341} (1995) 325.
%
\bibitem{bj} 
J.~D.~Bjorken, Phys. Rev. {\bf D27} (1983) 140. 
\bibitem{LeBellac2}
M.~Le~Bellac and H.~Mabilat, preprint INLN 96/17, July 1996.
%
\bibitem{Bedaque}
For a discussion of pinch singularities, which differs from
 \cite{Altherr2}, \\
P.~F.~Bedaque, Phys. Lett. {\bf B344} (1995) 23. 
%
\end{thebibliography}
\end{document}